\documentclass[a4paper,fleqn]{jaa}
\usepackage{graphicx}	% Including figure files
\usepackage{natbib,hyperref}
\usepackage{color,rotating}
\usepackage{amsmath}	% Advanced maths commands
\usepackage{amssymb}	% Extra maths symbols
\usepackage{multicol}	% Multi-column entries in tables
\usepackage{bm}		% Bold maths symbols, including upright Greek
\usepackage{pdflscape}	% Landscape pages
\usepackage{multirow}
\def\prd{Phys. Rev. D}

\def\aap{A \& A}
\def\apjs{ApJS}
\def\mnras{MNRAS}
\def\apj{ApJ}
\def\apjl{ApJL}

\def\nat{Nature}

\def\aj{The Astronomical Journal}
\def\pasa{Publications of the Astronomical Society of Australia}

\def\ud{\mathrm{d}}

\def\HI{H~{\sc i}\,}
\def\HII{H~{\sc ii}\,}
\def\xb{\bar{x}_{{\rm H{\sc{I}}}}}

\def\Pl{{\mathcal P}_l}

\def\rhohi{\rho_{H{\sc I}}}
\def\rhoh{\rho_{\mathrm{M}}}
\begin{document}
\title[LoS anisotropies in the CD-EoR 21-cm power spectrum]{Line of
  sight anisotropies in the Cosmic Dawn and EoR 21-cm power spectrum}
%
%--------------------------------------------------------------------
\author[Majumdar et al.]{Suman Majumdar$^1$\thanks{Email:
    s.majumdar@imperial.ac.uk}, Kanan K. Datta$^2$, Raghunath Ghara$^3$,
  \newauthor Rajesh Mondal$^4$, T. Roy Choudhury$^3$, Somnath Bharadwaj$^4$,\newauthor Sk. Saiyad Ali$^5$ \& Abhirup Datta$^6$ \\
  $^1$Department of Physics, Blackett Laboratory, Imperial College, London SW7 2AZ, UK \\
  $^2$Department of Physics, Presidency University, 86/1 College Street, Kolkata - 700073, India\\
  $^3$National Centre for Radio Astrophysics, TIFR, Post Bag 3,
  Ganeshkhind, Pune 411007, India\\ $^4$Department of Physics \&
  Centre for Theoretical Studies, Indian Institute of Technology
  Kharagpur,\\ Kharagpur - 721302, India\\ $^5$Department of
  Physics, Jadavpur University, Kolkata 700032, India\\ $^6$Centre for
  Astronomy, Indian Institute of Technology Indore, Indore - 452020,
  India}

\date{}
%\date{\today}
%\pubyear{2016}

%--------------------------------------------------------------------

\label{firstpage}
\pagerange{\pageref{firstpage}--\pageref{lastpage}}
\maketitle

%--------------------------------------------------------------------

\begin{abstract}
The line of sight direction in the redshifted 21-cm signal coming from
the cosmic dawn and the epoch of reionization is quite unique in many
ways compared to any other cosmological signal. Different unique
effects, such as the evolution history of the signal, non-linear
peculiar velocities of the matter etc will imprint their signature
along the line of sight axis of the observed signal. One of the major
goals of the future SKA-LOW radio interferometer is to observe the
cosmic dawn and the epoch of reionization through this 21-cm
signal. It is thus important to understand how these various effects
affect the signal for it's actual detection and proper
interpretation. For more than one and half decades, various groups in
India have been actively trying to understand and quantify the
different line of sight effects that are present in this signal
through analytical models and simulations. In many ways the importance
of this sub-field under 21-cm cosmology have been identified,
highlighted and pushed forward by the Indian community. In this
article we briefly describe their contribution and implication of
these effects in the context of the future surveys of the cosmic dawn
and the epoch of reionization that will be conducted by the SKA-LOW.

\end{abstract}

%--------------------------------------------------------------------


%--------------------------------------------------------------------

%\input{intro}
%\input{los_effects}
%\input{method_anisotropy}
%\input{quantify_anisotropy}
%\input{detectability}
%\input{summary}

%--------------------------------------------------------------------
\section{Introduction}
\label{sec:intro}
The `Cosmic Dawn' (CD) is the period in the history of our universe
when the first sources of light were formed and they gradually warmed
up their surrounding intergalactic medium (IGM). This period was
followed by the `Epoch of Reionization' (EoR) when these first and
subsequent population of sources produced enough ionizing photons to
gradually change the state of most of the hydrogen in our universe
from neutral (\HI\!\!)  to ionized (\HII\!\!). These two epochs
possibly are the least known periods in the history of our
universe. Our present understanding of these epochs is mainly
constrained by the observations of the cosmic microwave background
radiation (CMBR) \citep{komatsu11, planck15},the absorption spectra of
the high redshift quasars \citep{becker01,fan03,becker15} and the
luminosity function and clustering properties of Ly$\alpha$ emitters
(see e.g. \citealt{trenti10,ouchi10,bouwens16}). These indirect
observations are rather limited in their ability to answer many
fundamental questions regarding these periods; {\it e.g.} the
properties of these first sources of light and how they evolve as
reionization progresses, precise duration and timing of the CD and the
EoR, the relative contribution in the heating and ionization of the
IGM from various types of sources, and the typical size and
distribution of the heated and ionized regions. It is being
anticipated that the currently ongoing and proposed future radio
interferometric surveys of these epochs, through the brightness
temperature fluctuations of the redshifted 21-cm signal originating
due to the hyperfine transition in the neutral hydrogen, has the
potential to answer most these fundamental questions.

The currently ongoing redshifted 21-cm radio interferometric surveys,
being conducted using the
{GMRT\footnote{{http://www.gmrt.ncra.tifr.res.in}}}
\citep{paciga13},
{LOFAR\footnote{{http://www.lofar.org}}}
\citep{yatawatta13,haarlem13},
{MWA\footnote{{http://www.haystack.mit.edu/ast/arrays/mwa}}
  \citep{tingay13,bowman13,dillon14}} and
{PAPER\footnote{{http://eor.berkeley.edu}}}
\citep{parsons14,jacobs14,ali15}, are mainly aimed at detecting this
signal from the EoR. Most of these instruments are not sensitive
enough at high redshifts to detect this signal from the CD. The low
frequency aperture array of the upcoming Square Kilometre Array
({SKA-LOW\footnote{{http://www.skatelescope.org}}})
will be a significant step forward in this regard as it will have
enough sensitivity over a large range of low frequencies to observe
the redshifted 21-cm signal from both the CD and the EoR
\citep{mellema13,koopmans15}. Also, it is worth mentioning that, while
the first generation experiments aim to detect the signal through
statistical estimators like variance and power spectrum (due to their
limited sensitivity), the SKA-LOW is expected to be sensitive enough
(owing to its large collecting area) to make images of fluctuations in
\HI from these epochs \citep{mellema15}.

However, as the redshifted 21-cm signal from the CD and the EoR has
not been detected till date by any of these first generation
telescopes and they so far only managed to provide somewhat weak upper
limits on the signal power spectra \citep{paciga13,ali15} at large
length scales, so it has been planned that the first phase of the
SKA-LOW will survey a large volume of the sky for a relatively shorter
observation time to estimate the signal power spectrum at different
redshifts, which will possibly constrain some of the main parameters
of the signal \citep{koopmans15}. The spherically averaged power
spectrum which is perceived as the main tool to achieve the first
detection of the signal with the first generation instruments and the
initial tool for the CD-EoR parameter estimation with the SKA-LOW,
provides a high signal to noise ratio (SNR) by averaging the signal in
spherical shells in Fourier space, while still preserving many
important features of the signal. However, as the CD-EoR 21-cm signal
is expected to be highly non-Gaussian in nature
\citep{bharadwaj05a,mondal15,mondal16,mondal16b}, thus the power
spectrum alone cannot represent all the properties of such a field.

Even when one is dealing with the power spectra of the signal, one has
to be aware of the fact that the line of sight (LoS) direction of the
redshifted 21-cm signal is unique. As the 21-cm signal originates from
a line transition, signal coming from different cosmological distances
along the LoS essentially belongs to different epochs and gets
redshifted to different wavelengths, characterized by their
cosmological redshift $z$. This implies that the signal present in an
actual observational data cube containing a range of frequencies or
wavelengths will evolve in time along the frequency or LoS
direction. This is popularly known as the light cone effect
\citep{barkana06,datta12,datta14}. Thus while analyzing the three
dimensional data one has to take this into account for a proper
interpretation of the signal.

Another effect which also affects the signal along the LoS direction
is the non-random distortions of the signal caused by the peculiar
velocities of the matter particles. The coherent inflows of matter
into overdense regions and the outflows of matter from underdense
regions will produce an additional red or blueshift in the 21-cm
signal on top of the cosmological redshift, changing the contrast of
the 21-cm signal, and making it anisotropic along the LoS. It has been
first highlighted by \citet{bharadwaj04,bharadwaj05} and \citet{ali05}
that the peculiar velocities will significantly change the amplitude
and the shape of the 21-cm power spectrum measured from the
observations of the periods before, during and even after the
reionization.

\citet{ali05} was also first to highlight another source of anisotropy
along the LoS of the redshifted 21-cm signal from the CD and the EoR,
which is caused due to the Alcock-Paczynski effect
\citep{alcock79}. This is the anisotropy due to the non-Euclidean
geometry of the space-time. It is important to take into account this
effect when one wants to do parameter estimation using the CD-EoR
21-cm power spectrum.

The `bulk flows', which could arise from possible supersonic relative
velocities between dark matter and baryonic gas \citep{tselia10} can
also contribute to the LoS anisotropy in the signal.

Understanding and quantifying the line of sight anisotropy (caused by
several aforementioned reasons) in the CD-EoR 21-cm power spectrum has
been one of the long standing important issues in 21-cm
cosmology. This will be especially important in the context of the
upcoming SKA-LOW observations, which will probably be the first
instrument to have enough sensitivity to constrain the model
parameters of both the CD and the EoR significantly using the 21-cm
power spectrum. We have been very active for almost one and half decade 
in this sub-field under 21-cm cosmology. To be more precise, in the 
context of the 21-cm cosmology this sub-field was identified and 
highlighted by \citet{bharadwaj01,bharadwaj04,bharadwaj05,bharadwaj05a,ali05,datta07a}, through analytical models of the signal power spectrum. These
analytical predictions were later tested, validated and further pushed
forward by the same groups using various kinds of simulations
\citep{choudhury09b,datta12,majumdar13,jensen13,datta14,majumdar14,ghara14,ghara15,majumdar16}. In this article we aim to summarize the effects 
of line of sight anisotropy in the CD-EoR 21-cm power spectrum 
considering future SKA-LOW observations and our contribution in 
this subject.

The structure of this article is as follows. In Section
\ref{sec:los_effects}, we briefly describe different sources that
contribute to the LoS anisotropy of the signal. We next describe
different methods to quantify the LoS anisotropy present in the power
spectra in Section \ref{sec:methods}.  In Sections
\ref{sec:quantifying_rsd} and \ref{sec:quantifying_lc} we describe how
one would be able to quantify and interpret the two major sources of
LoS anisotropy in the signal using the simulated expected signal at
different era, for several ongoing experiments and as well as for the
SKA-LOW. We further discuss several issues that may hinder the
detection of the signal as well as the quantification of the
anisotropy in it in Section \ref{sec:detectability}. Finally, in
Section \ref{sec:summary}, we summarize our review.

%Throughout this paper, we have used the Planck+WP best fit values of
%cosmological parameters $\Omega_{\rm m0}=0.3183$,
%$\Omega_{\rm \Lambda0}=0.6817$, $\Omega_{\rm b0}h^2=0.022032$, $h=0.6704$,
%$\sigma_8=0.8347$, and $n_{\rm s}=0.9619$ \citep{planck14}.
%-------------------------------------------------------------------------
\section{Line of sight anisotropies in the redshifted 21-cm signal from the CD and EoR}
\label{sec:los_effects}
We briefly describe here the three major LoS anisotropies affecting
the redshifted 21-cm signal originating from the cosmic dawn and the
EoR. 
\subsection{Redshift space distortions}
The fluctuations in the brightness temperature of the redshifted 21-cm
radiation essentially trace the \HI distribution during the CD and the
EoR. In a completely neutral IGM the \HI distribution is expected to
follow the underlying matter distribution with a certain amount of
bias, at large enough length scales. The coherent inflows of matter
into overdense regions and the outflows of matter from underdense
regions will produce an additional red or blue shift on top of the
cosmological redshift. If we consider a distant observer located along
the $x$ axis and use the $x$ component of the peculiar velocity to
determine the position of \HI particles in redshift space
\begin{equation}
s = x + \frac{v_x}{a H(a)}\,,
\end{equation}
where $a$ and $H(a)$ are the scale factor and Hubble parameter
respectively. Thus, in redshift space the apparent locations of the
\HI particles will change according to the above equation, which will
effectively change the contrast of the 21-cm signal and will make it
anisotropic along the LoS. However, as the first sources of lights
were formed and they start converting their surrounding \HI into \HII
, the one-to-one correspondence between matter and \HI no longer holds
and the effect of the matter peculiar velocities on the 21-cm signal
becomes much more complicated. In the context of the 21-cm signal from
the CD and EoR, this effect was first pointed out and quantified in
\citet{bharadwaj04} and \citet{bharadwaj05}. Their analytical
treatment showed that the redshift space distortions changes the shape
and amplitude of the signal power spectrum significantly when measured
from the recorded visibilities in a radio interferometric observation.
It has also been proposed that one can possibly extract the matter
power spectrum \citep{barkana05,shapiro13} at these epochs using the
redshift space anisotropy present in the 21-cm power spectrum.

\subsection{Light cone effect}
The other LoS anisotropy that will be present in the observed 21-cm
signal is known as the light cone anisotropy. As light takes a finite
amount of time to reach from a distant point to an observer, thus the
cosmological 21-cm signal coming from different cosmological redshifts
essentially belongs to different distances and thus correspond to
different cosmological epochs. In the context of redshifted 21-cm
signal, this change in frequency due to cosmological redshift can be
represented by
\begin{equation}
  \lambda_{\rm obs} = \lambda_{\rm emitted}(1+z) \,.
\end{equation}
Any radio interferometric observation produces a three dimensional
data set containing a range of frequencies. The time evolution of the
signal will be present in such a data set as one changes the
frequency. Thus, while estimating power spectrum of the signal from
such a data one needs take into account this effect for a proper
interpretation of the signal. In the context of 21-cm signal this was
first considered by \citet{barkana06} in their analytical estimation
of the two point correlation function of the signal. This also makes
the shape of the \HII regions around extremely bright sources
(e.g. quasars) anisotropic along the LoS
\citep{wyithe05,yu05,majumdar11,majumdar12}. If the reionization was
dominated by such sources than the effect on power spectrum and
correlation functions in such a case has been studied by
\citet{sethi08}. Such anisotropy in the shape of the \HII region can
be detected by targeted tomographic imaging by deep observations
around such sources using of the SKA-LOW \citep{majumdar12}. This has
been discussed in further details in a different review article in
this special issue.

\subsection{Alcock-Paczynski effect}
The Alcock-Paczynski effect (\citealt{alcock79}, hereafter the AP
effect), another anisotropy in the signal along the LoS, is caused due
to the non-Euclidean geometry of the space-time. This makes any
object, which is intrinsically spherical in shape, to appear elongated
along the LoS. At low redshifts (for $z \leq 0.1$) it is not that
significant but at high redshifts this causes a significant distortion
in the signal, which makes the power spectrum of the 21-cm signal from
the CD and the EoR anisotropic along the LoS. The AP effect in the
context of 21-cm signal from the EoR was first considered by
\citet{ali05} in their analysis of the signal through the power
spectrum. The proposal of \citet{barkana05}, that one can probably
distinguish different sources that contribute to the 21-cm power
spectrum by measuring the anisotropy in the power spectrum, has the
implicit assumption that the background cosmological model is known to
a great degree and hence does not take into account the anisotropies
introduced by the geometry (AP effect). \citet{ali05} in contrast
adopted a framework which allows the high redshift 21-cm signal to be
interpreted without reference to a specific background cosmological
model and also studied the variation in the anisotropy due to the AP
effect depending on different background cosmological models. They
further quantify the relative contribution in anisotropy due to the AP
effect when compared with the anisotropy due to the redshift space
distortions and how they differ in their nature. However, it has not
been studied yet, how significant this effect will be when compared to
the uncertainties due to the thermal noise and other systematics for a
future SKA-LOW observation.
%------------------------------------------------------------------------
\section{Methods to quantify the LoS anisotropy in 21-cm signal}
\label{sec:methods}
We next discuss different possible ways of quantifying any line of
sight anisotropy present in the signal through its power spectrum.

\subsection{$\mu$-decomposition of the power spectrum}
It is convenient to introduce a parameter $\mu$, defined as the cosine
of the angle between a specific Fourier mode $\mathbf{k}$ and the
LoS. In case of plane parallel redshift space distortions, the
redshift space power spectrum then can be expressed as a fourth-order
polynomial in $\mu$ (see e.g. \citealt{bharadwaj04,bharadwaj05,
  ali05,barkana05}):
\begin{align}
\label{eq:rsd_mu}
P^{\mathrm{s}}(k, \mu) = \overline{\delta T_b}^2(z) \left[
  P_{\mu^0}(k) + \mu^2 P_{\mu^2}(k) + \mu^4 P_{\mu^4}(k) \right]\,,
\end{align}
where $\overline{\delta T_b}$ is the average differential brightness
temperature of the 21-cm signal at a specific redshift $z$. In the
context of the CD and EoR 21-cm signal this representation of the
power spectrum is popular, as one can directly identify each
coefficients of the powers of $\mu$ with physical quantities
contributing to the 21-cm power spectrum under the linear
\citep{bharadwaj01,bharadwaj04,bharadwaj05, ali05,barkana05,
  lidz08,majumdar13} or quasi-linear
\citep{mao12,jensen13,majumdar14,majumdar16,ghara14,ghara15} model for
the signal (which is valid mainly for the large scale fluctuations in
the signal). Following this quasi-linear model one can express each of
these coefficients of $\mu$ as:
\begin{align}
\label{eq:rsd_qlin}
\nonumber P_{\mu^0}(k)  &= P_{\rhohi,\rhohi}(k) + P_{\eta,\eta}(k) + P_{\rhohi,\eta}(k) \,, \\ 
\nonumber P_{\mu^2}(k)  &= 2  \left[P_{\rhohi, \rhoh}(k) + P_{\rhoh, \eta}(k) \right]\,, \\
 P_{\mu^4}(k)  &= P_{\rhoh,\rhoh}(k) \,.
\end{align}
where $\rhohi$ is the neutral hydrogen density, $\rhoh$ is the total
hydrogen density and $\eta (z,\mathbf{x})= 1 - T_{\mathrm
  CMB}(z)/T_{\mathrm S}(z,\mathbf{x})$ represents spin temperature
fluctuations in the \HI distribution. The spin temperature
($T_{\mathrm S}$), which represents the relative population of atoms
in two different spin states, can get affected by the Lyman-$\alpha$
pumping and heating during the Cosmic Dawn and the early stages of
reionization
\citep{bharadwaj01,bharadwaj04,bharadwaj05,ali05,ghara14,ghara15}. For
most part of the reionization, by when the IGM is expected to be
significantly heated above the CMB temperature, it can be safely
assumed that $T_{\mathrm S} \gg T_{\mathrm CMB}$ (unless it is a case
of very late heating). In that case, all terms related to $\eta$ turns
out to be zero and it becomes very tempting to conclude that at least
at large scales this representation will hold and it would be possible
to separate the astrophysics ($P_{\rhohi,\rhohi}$ and $P_{\rhohi,
  \rhoh}$) from cosmology ($P_{\rhoh,\rhoh}$). However, one has to be
aware of the fact that this representation is not in orthonormal
basis, thus each of these coefficients are not independent of the
other, thus decomposing the power spectrum in this form and
identifying the coefficients of the powers of $\mu$ with certain
physical quantities may lead to erroneous conclusions. It is also
important to note that this model do not consider any AP effect to be
present in the signal, which will introduce an additional $\mu^6$ term
in eq. \eqref{eq:rsd_mu} \citep{ali05}.
\subsection{Legendre polynomial decomposition of the power spectrum}
A different approach to quantify the LoS anisotropy is to instead
expand the power spectrum in the orthonormal basis of Legendre
polynomials, which is a well known approach in the field of galaxy
redshift surveys \citep{hamilton92,hamilton98,cole95}. In this
representation (assuming plane parallel redshift space distortions), the
power spectrum can be expressed as a sum of the even multipoles of
Legendre polynomials:
\begin{equation}
  P^{\mathrm{s}}(k,\mu) = \sum_{l\;\mathrm{even}} \Pl(\mu) P_l^{\mathrm{s}}(k).
  \label{eq:legendre}
\end{equation}
From an observed or simulated 21-cm power spectrum,
one can calculate the angular multipoles $P_l^{\mathrm{s}}$ following:
\begin{equation}
  P_l^{\mathrm{s}}(k) = \frac{2l + 1}{4\pi} \int \Pl(\mu) P^{\mathrm{s}}(k) \ud \Omega .
  \label{eq:multipole}
\end{equation}
where $\Pl(\mu)$ represents different Legendre polynomials. The integral
is performed over the entire solid angle to take into account all
possible orientations of ${\bf k}$ with the LoS direction. Each
multipole moment estimated through eq.  \eqref{eq:multipole} will be
independent of the other, as this is a representation in orthonormal
basis. In the context of 21-cm signal from the EoR, the effect of
redshift space distortions was first quantified in \citet{majumdar13}
by estimating the quadrupole ($P_2^{\mathrm{s}}$) and monopole
($P_0^{\mathrm{s}}$) moments of the power spectrum from EoR
simulations. If one considers the quasi-linear model of the signal,
the first three non-zero angular multipoles of the power spectrum can
be expressed as \citep{majumdar14,majumdar16}:
\begin{align}
  P^s_0 &= \overline{\delta T_b }^2(z) \left[ \frac{1}{5} P_{\rhoh,
      \rhoh} + P_{\rhohi, \rhohi} + P_{\eta, \eta} +   2 P_{\eta, \rhohi} \right. \label{eq:eta_P0}\\ \nonumber
  &\left. + \frac{2}{3} P_{\rhohi, \rhoh} + \frac{2}{3} P_{\eta, \rhoh} \right]  \\
  P^s_2 &= 4\, \overline{\delta T_b }^2(z) \left[ \frac{1}{7} P_{\rhoh, \rhoh}  + \frac{1}{3} P_{\rhohi, \rhoh} + \frac{1}{3} P_{\eta, \rhoh} \right] \label{eq:eta_P2}\\
  P^s_4 &= \frac{8}{35}\, \overline{\delta T_b }^2(z)\, P_{\rhoh,
    \rhoh} \label{eq:eta_P4}
\end{align}
This shows that it might be possible to extract the matter power
spectrum by estimating the hexadecapole moment ($P_4^{\mathrm{s}}$) of
21-cm power spectrum for sufficiently large length scales. However, at
those scales the signal might be severely dominated by the cosmic
variance (see Section \ref{subsec:cosmic_variance} for a detailed
discussion). It further shows that, it is not possible to
independently extract $P_{\rhohi,\rhohi}$ or $P_{\rhohi, \rhoh}$, even
when $T_{\mathrm S} \gg T_{\mathrm CMB}$ ({\it i.e.} when all $\eta$
related terms are zero).

\subsection{Other alternative method}
A simpler alternative method to quantify the LoS anisotropy in the
signal power spectrum is to calculate the anisotropy ratio proposed by
\citet{fialkov15}, which is defined as:
\begin{equation}
\label{eq:r_mu}
r_{\mu}(k, z) = \frac{\langle P({\mathbf k}, z)_{|\mu_k|>0.5}\rangle}{\langle P({\mathbf k}, z)_{|\mu_k|<0.5}\rangle} - 1\,,
\end{equation}
where the angular bracket represent average over angles. Ideally, if
the signal is isotropic then $r_{\mu}(k, z)$ should be zero, otherwise
it will take positive or negative values. The redshift evolution of
this quantity will some way quantify the integrated or angle averaged
degree of anisotropy present in the signal at different stages of the
CD and the EoR. 
%----------------------------------------------------------------------

\section{Quantifying the redshift space distortions in the 21-cm power
  spectrum}
\label{sec:quantifying_rsd}
\subsection{During the EoR}
\begin{figure*}
%\psfrag{P2/P0}[c][c][1][0]{\large{$P^s_2/P^s_0$}}
%\psfrag{xh1}[c][c][1][0]{\large{$\xb$}}
%\psfrag{k = 0.23 Mpc-1}[c][c][1][0]{\tiny{$k = 0.23 \,{\rm Mpc}^{-1}$}}
%\psfrag{k = 0.059 Mpc-1}[c][c][1][0]{\tiny{$k = 0.06 \,{\rm Mpc}^{-1}$}}
%\psfrag{k = 0.12 Mpc-1}[c][c][1][0]{\tiny{$k = 0.12 \,{\rm Mpc}^{-1}$}}
%\psfrag{C2RAY}[r][r][1][0]{\tiny{RT}}
%\psfrag{Sem-Num e=0.0}[r][r][1][0]{\tiny{Sem-Num}}
%\psfrag{21cmFASTL}[r][r][1][0]{\tiny{CPS+GS}}
\includegraphics[width=1.0\textwidth,angle=0]{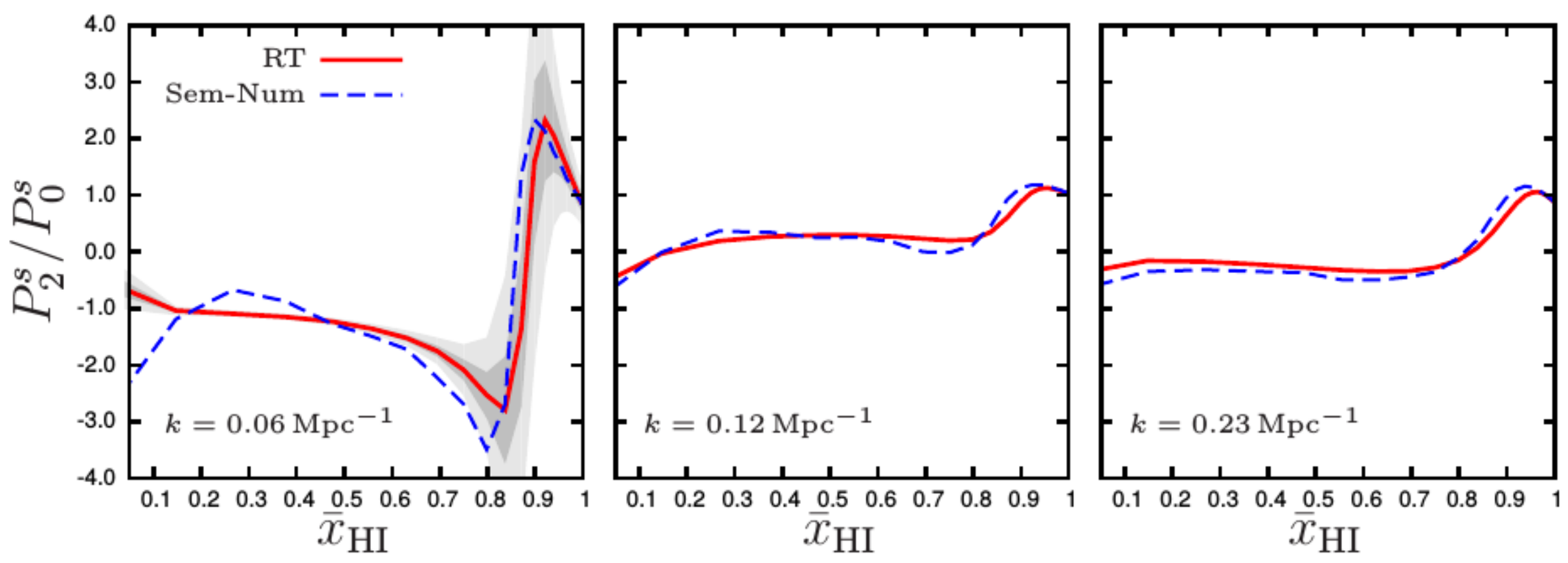}

\caption{The evolution of the ratio $P^s_2/P^s_0$ with $\xb$ at three
  representative $k$ values estimated from a set of inside-out
  simulations of the EoR 21-cm signal (one semi-numerical and the
  other radiative transfer). The shaded regions in
  light and dark gray represent uncertainty due to the system noise
  for $2000$ and $5000$ hr of observation using a LOFAR like
  instrument at $150$ MHz. Figure taken from \citet{majumdar14}.}
\label{fig:P2P0_speck}
\end{figure*} 
The anisotropy in the redshifted 21-cm signal from the EoR has been
traditionally quantified through the ratio of the spherically averaged
power spectrum in redshift space to the same quantity estimated in
real space. Predictions for this ratio has been made from analytical
models as well as from radiative transfer and semi-numerical
simulations of the signal ({\it e.g.} \citealt{lidz07,
  mesinger11,mao12} etc.). However, one very important point to note
here that, in reality one would not be able to estimate this ratio, as
the signal will always have redshift space distortions present in it,
thus it is not possible to estimate the power spectrum of the signal
in real space. Instead of estimating this ratio, \citet{majumdar13}
suggested that one could independently estimate different angular
multipole moments of the power spectrum of the signal (following
eq. [\ref{eq:legendre}] and [\ref{eq:multipole}]) directly from the
observed visibilities, which is the basic observable in any radio
interferometric survey. They further suggested that, the ratio between
the first two non-zero even multipole moments of the power spectrum
({\it i.e.}  monopole or $P^s_0$ and quadrupole or $P^s_2$) can be
used as an estimator to quantify any LoS anisotropy present in the
signal. It is precisely due to the fact that, if there is no LoS
anisotropy in the signal, any higher order multipole moment, other
than the monopole moment (which by definition is the spherically
averaged power spectrum) will be zero. To quantify the effect of
redshift space distortions in the signal \citet{majumdar13} used a set
of semi-numerical simulations (with a proper implementation of the
redshift space distortions in it using actual matter peculiar
velocities) of the signal and studied the evolution of the ratio of
$P^s_2/P^s_0$ with an evolving global neutral fraction ($\xb$) at
different length scales observable to the present and future EoR
surveys. They observed that at the early stages of the EoR ($1\gtrsim
\xb \gtrsim 0.9$), for an inside-out reionization scenario, at large
and intermediate length scales ($0.5 \gtrsim k \gtrsim 0.05\, {\rm
  Mpc}^{-1}$), this ratio is positive and gradually goes up from the
initial value of around $P^s_2/P^s_0 = 49/50$ (predicted by the linear
model for a completely neutral IGM) to higher values ($\approx 3$ for
$k \approx 0.05\, {\rm Mpc}^{-1}$ at $\xb \approx 0.9$) as
reionization progresses further. Once the early stage of the EoR is
over, one observes a sharp transition in this ratio at around $\xb
\approx 0.9$, where it becomes negative and reaches values as low as
$\approx -3$ at large length scales. It slowly goes up again with
decreasing neutral fraction (for $0.5 \lesssim \xb \lesssim 0.9$) and
reaches a value of $-1$ by $\xb \approx 0.5$. This ratio remains
negative and almost constant ($P^s_2/P^s_0 \approx -1$) for the rest
of the reionization ($\xb \leq 0.5$), until the signal strength goes
down and it becomes undetectable. \citet{majumdar13} ascribes the
sharp peak and dip features of the $P^s_2/P^s_0$ versus $\xb$ curve
around $\xb \approx 0.9$ and negative value of $P^s_2/P^s_0$ for $\xb
\leq 0.9$ to the inside-out nature of the reionization. The robustness
of these results (and associated features in the behaviour of
$P^s_2/P^s_0$) were further confirmed by \citet{majumdar14} by
estimating the same ratio from a radiative transfer and a set of
semi-numerical simulations (Figure \ref{fig:P2P0_speck}). They also
showed that this ratio would be detectable using LOFAR after $2000$ hr
of observations. As these features of this ratio appears to be model
invariant, so it can be used as a confirmative test for the detection
of the signal. Both of the above mentioned studies also find that the
hexadecapole moment ($P^s_4$) will be severely dominated by cosmic
variance and thus it would be difficult to conclude anything about the
matter power spectrum through it.

In a similar study with a radiative transfer simulation, but using the
$\mu$-decomposition technique described in Section \ref{sec:methods}
through eq. \eqref{eq:rsd_mu}, \citet{jensen13} also found that it is
not possible to extract the coefficients of $\mu^2$ and $\mu^4$ terms
in the signal power spectrum, {\it i.e.} not possible to extract
$P_{\rhohi, \rhoh}$ and $P_{\rhoh,\rhoh}$ independently, rather it is
possible to extract the sum of the coefficients of $\mu^2$ and
$\mu^4$, though it to would be more prone to errors compared to
estimating the quadrupole moment ($P^s_2$) of the power spectrum, as
it is a representation in non-orthonormal basis.

\subsubsection{Constraining the EoR history using the redshift space
  anisotropy in the 21-cm signal}
\begin{figure*}
%\psfrag{xh1}[c][c][1][0]{\Large{$\xb$}}
%\psfrag{P2}[c][c][1][0]{\Large{$k^3P^s_{2}/(2\pi^2)\, ({\rm mK^2})$}}
%\psfrag{k = 0.122 Mpc-1}[c][c][1][0]{{$k=0.12\,{\rm Mpc}^{-1}$}}
%\psfrag{fidu}[c][c][1][0]{Fiducial$\,\,\,\,\,\,\,\,$}
%\psfrag{fidu inh}[c][c][1][0]{{Clumping$\,\,\,\,$}}
%\psfrag{xray=0.8}[c][c][1][0]{{UIB Dom$\,$}}
%\psfrag{xrays=0.8}[c][c][1][0]{{SXR Dom}}
%\psfrag{xraysh=0.5}[c][c][1][0]{{UV+SXR+UIB$\,\,\,\,\,\,\,\,\,\,\,$}}
%\psfrag{pl=2.0}[c][c][1][0]{{PL 2.0}}
%\psfrag{pl=3.0}[c][c][1][0]{{PL 3.0}}
\includegraphics[width=1.0\textwidth,angle=0]{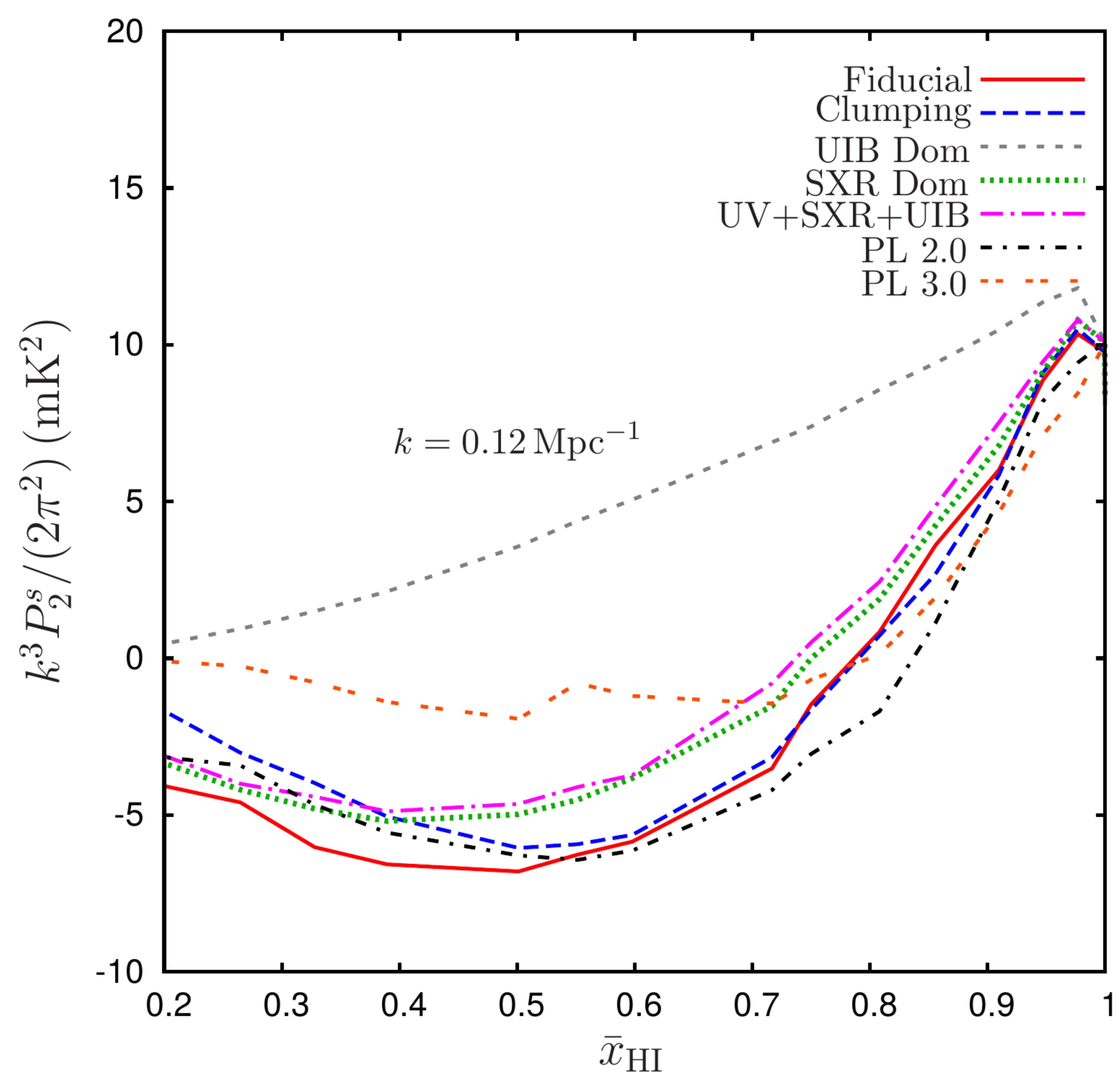}
\caption{The quadrupole moment of the power spectrum for different
  reionization scenarios as a function of the global neutral fraction
  at $k=0.12$ Mpc$^{-1}$. Figure taken from \citet{majumdar16}.}
\label{fig:p2}
\end{figure*}
The quasi-linear model in eq. \eqref{eq:eta_P2} suggest that the
quadrupole moment ($P^s_2$) contains matter power spectrum and the
cross-power spectrum between the matter and \HI (when $T_{\mathrm S}
\gg T_{\mathrm CMB}$, {\it i.e.} all $\eta$ related terms are
zero). While the matter power spectrum contains the amplitude of the
matter density fluctuations (which evolves rather slowly with
redshift), the cross-power spectrum $P_{\rhohi, \rhoh}$ contains the
information of the phase difference between the matter and the \HI
field. If the distribution and the properties of the reionization
sources change, the topology of the ionization field will also change
and one would expect then the phase difference between the matter and
the \HI field to also change, which it turn should have a signature in
the quadrupole moment. To test whether this idea can be used to
distinguish between different reionization sources using the nature
and amplitude of $P^s_2$, \citet{majumdar16} simulated a collection of
reionization scenarios considering various degrees of contribution
from different kinds of reionization sources. The reionization
scenarios they considered include ionizing photon contribution from
the usual UV photon sources hosted by the halos with mass $\gtrsim
10^9\,{\rm M_{\odot}}$, a uniform ionizing background generated by
hard X-ray sources, a local uniform ionizing background generated by
soft X-ray sources (limited by the mean free path of soft X-ray
photons), various combinations of all of these three contributions,
reionization driven by quasar like very strong sources located around
the most massive halos etc. They find that, for all of their
reionization scenarios (except the one dominated by a uniform ionizing
background), the quadrupole moment at large length scales ($k=0.12$
Mpc$^{-1}$) evolves with $\xb$ in a rather robust manner (Figure
\ref{fig:p2}), even though the topology of the 21-cm signal look
significantly different in all of those scenarios. They further show
that as long as the major ionizing photon sources follow the
underlying matter distribution, the phase difference between the
matter and the \HI evolves with $\xb$ in almost similar fashion,
though the topology of the 21-cm signal can be drastically
different. This is why the quadrupole moment ($P^s_2$) evolves in a
robust fashion with $\xb$ in all of those scenarios. Building on this
idea, they further demonstrate that this robustness of $P^s_2$ can be
used to extract the reionization history to a great degree. They show
that for an instrument with the sensitivity of the first phase of the
SKA-LOW, it will require $100$ hr (per pointing) of observation over
an area of minimum $5 \times 5 \,{\rm deg}^2$ in the sky to constrain
the reionization history ($\xb$ versus $z$) very precisely, if
foregrounds have already been removed to a great degree.
\subsection{During the Cosmic Dawn}
\begin{figure*}
\includegraphics[width=.5\textwidth,angle=0]{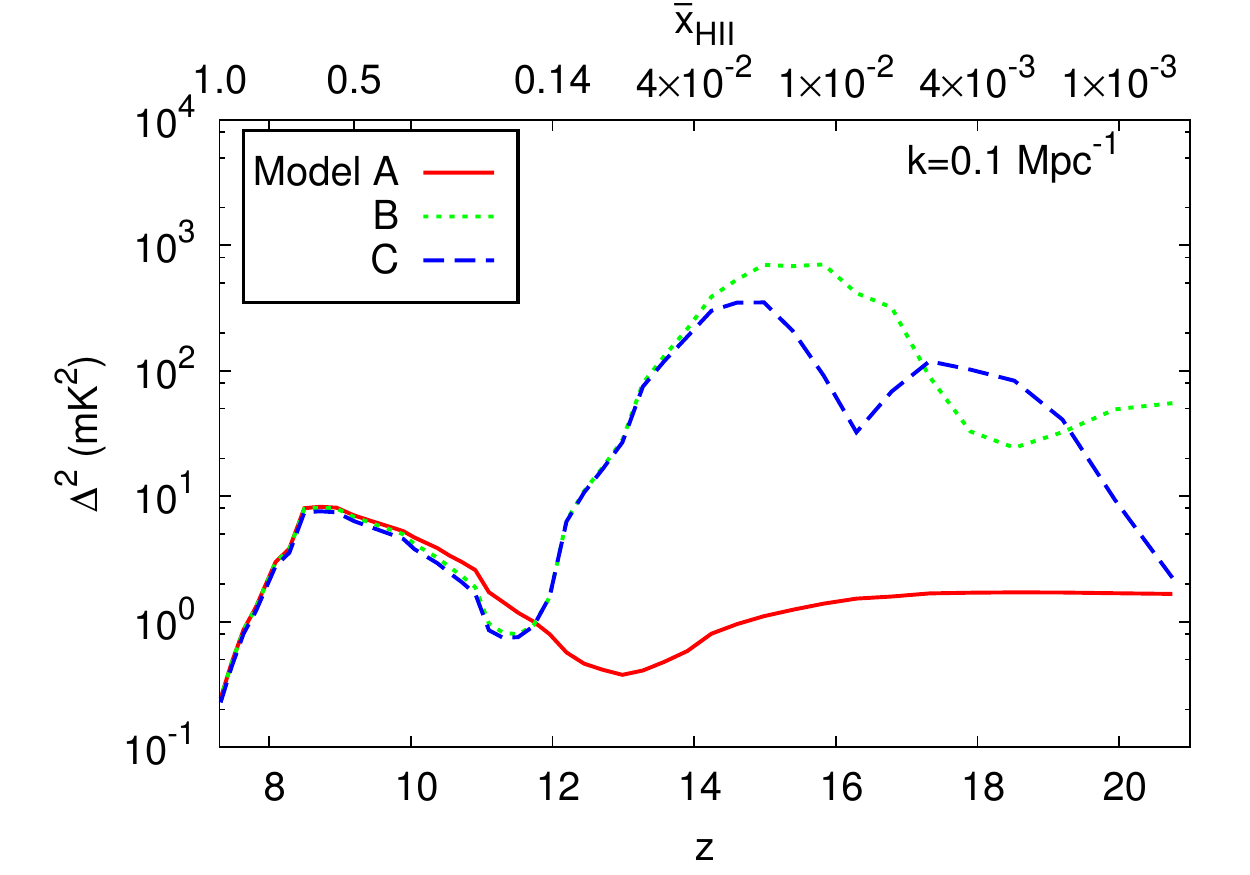}
\includegraphics[width=.5\textwidth,angle=0]{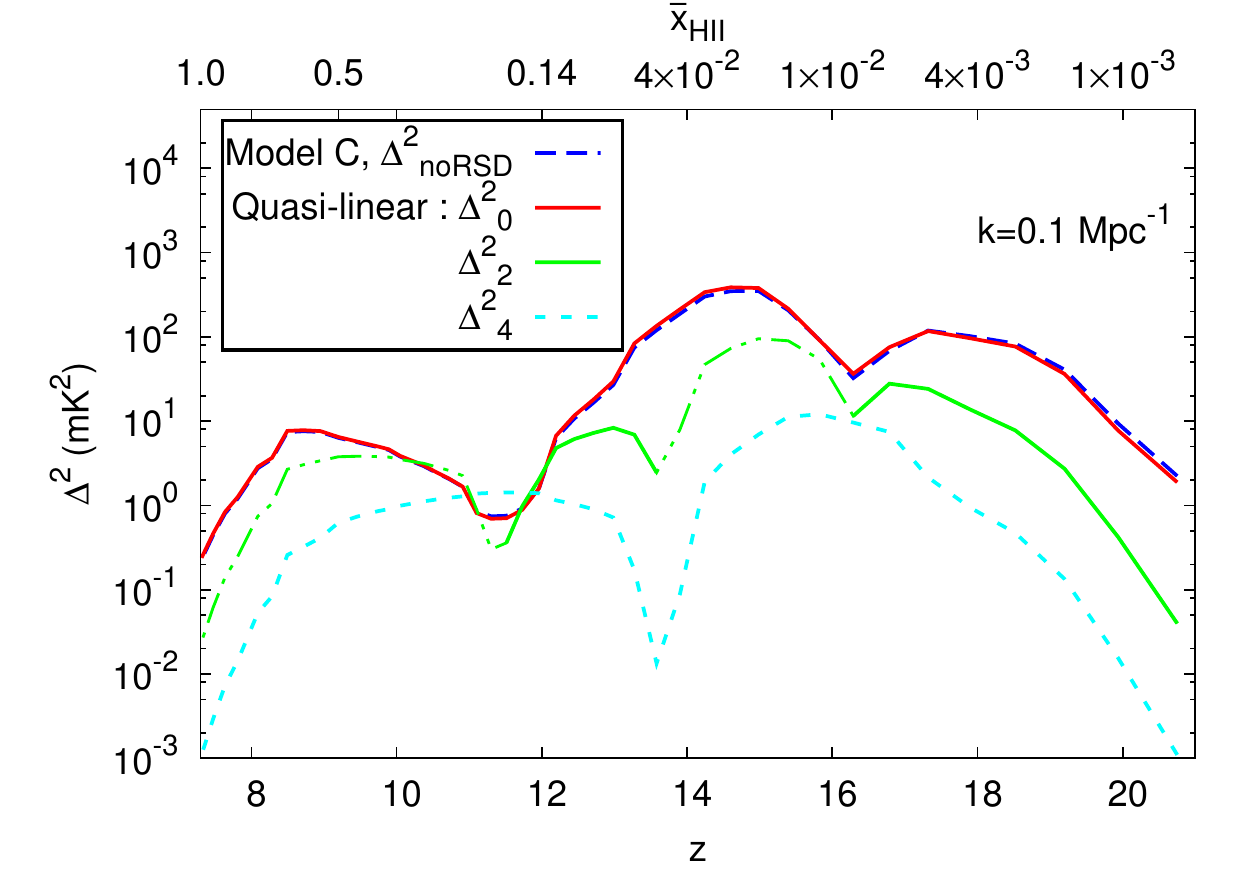}
\caption{The left panel shows the spherically averaged 21-cm power
  spectrum at large length scales for three different models of X-ray
  heating and Lyman-$\alpha$ coupling. Model A: IGM is Lyman-$\alpha$
  coupled and highly heated, Model B: IGM is strongly Lyman-$\alpha$
  coupled but self consistently heated with X-ray sources, Model C:
  IGM is self consistently coupled with Lyman-$\alpha$ and heated by
  X-ray sources. The right panel shows the redshift space
  $\mu$-decomposed power spectra estimated following
  eq. \eqref{eq:rsd_mu} for Model C. Figure taken from
  \citet{ghara14}.}
\label{fig:ps_cd}
\end{figure*}

The 21-cm signal from the Cosmic Dawn and the very early stages of the
EoR ($\xb \gtrsim 0.95$) are expected to be affected by the spin
temperature fluctuations caused by the inhomogeneous X-ray heating and
Lyman-$\alpha$ coupling around the early sources of light. In this
regime one would not be able to assume $T_{\mathrm S} \gg T_{\mathrm
  CMB}$ and the fluctuations in the quantity $\eta (z,\mathbf{x})= 1 -
T_{\mathrm CMB}(z)/T_{\mathrm S}(z,\mathbf{x})$ will contribute to the
21-cm brightness temperature significantly as predicted by
\citet{bharadwaj01,bharadwaj04,bharadwaj05,ali05} etc. \citet{ghara14}
developed and used an one dimensionals radiative transfer simulation
to study the effects of spin temperature fluctuations on the 21-cm
power spectrum during these stages while implementing the effect of
redshift space distortions in the signal using the actual matter
peculiar velocities. It was assumed that each of the early sources of
light produced in their simulation has two components: i) usual
stellar component (modelled using general population synthesis
prescriptions) and ii) a mini-quasar component (assumed to have a
power-law spectrum). They found some distinct features in the large
scale power spectrum in this case when compared to the scenario where
inhomogeneities in the gas temperature and the Lyman-$\alpha$ coupling
are ignored. They observed three distinct peaks in the large scale
spherically averaged power spectrum of the signal (which has been
reported earlier in the literature e.g. see \citealt{pritchard08}) when plotted as a function of
redshift (left panel of Figure \ref{fig:ps_cd}). The peak which
appears latest in the history is associated with $\sim 50\%$ neutral
fraction of the IGM and the \HI fluctuations have the maximum
contribution to the power spectrum at this stage. The second peak is
related to the fluctuations in the heating pattern and appears when
$\sim 10\%$ of the volume is heated above $T_{\rm CMB}$. The third
peak, which corresponds to the earliest stages of the 21-cm history,
corresponds to the inhomogeneities in the Lyman-$\alpha$
coupling. Identification of these peaks (two of which were not
reported earlier in the literature) would be very important when one
would try to parametrize the CD and the EoR using the observed power
spectrum and variance of the 21-cm signal from the proposed shallow
survey ($10$ hr observation covering $10,000\, {\rm deg^{2}}$) using
the first phase of SKA-LOW \citep{koopmans15}.

\begin{figure*}
\includegraphics[width=1.0\textwidth,angle=0]{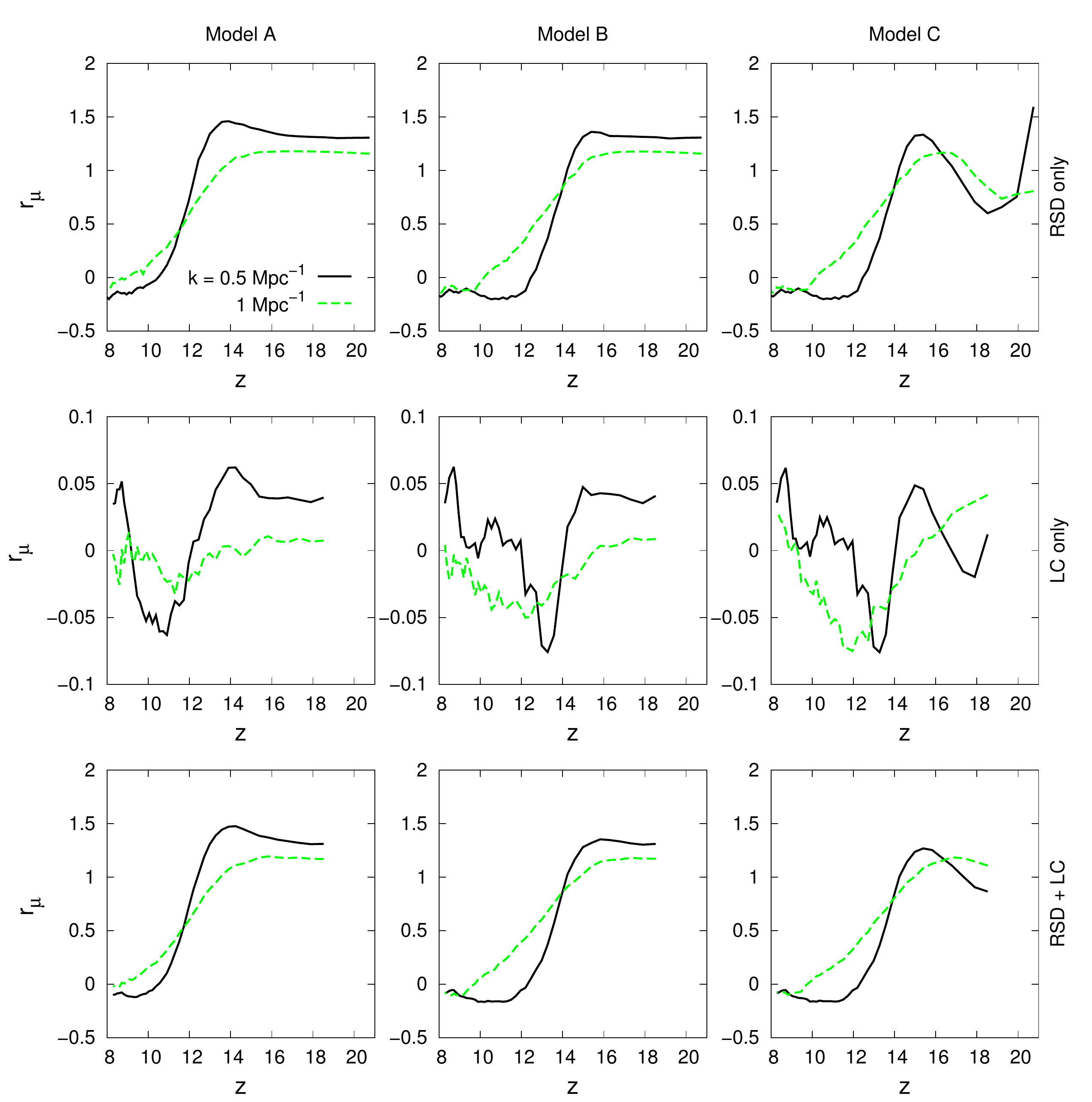}
\caption{This shows the anisotropy ratio $r_{\mu}$ computed following
  eq. \eqref{eq:r_mu} as a function of redshift for large and
  intermediate length scales. The models A, B and C are the same as
  described in Figure \ref{fig:ps_cd}, the only difference here is
  that all of them also include sources of mass lower than $10^9\,
  {\rm M_{\odot}}$. The three rows from top to bottom shows the ratio
  computed when the signal includes only the redshift space
  distortions, only light cone effect and both redshift space
  distortions and light cone effect, respectively. Figure taken from
  \citet{ghara15}.}
\label{fig:ani_r}
\end{figure*}
To quantify the effect of redshift space distortions \citet{ghara14}
estimated the ratio of the spherically averaged power spectrum in
redshift space and in real space. They find that this ratio at large
length scales to be not that high as reported by some earlier studies
\citep{majumdar13, jensen13}, when one includes the effect of spin
temperature fluctuations. They associate this disagreement to the fact
that they do not include the sources with mass lower than $10^9\, {\rm
  M_{\odot}}$ and also to the fact that during the CD and the early
stages of the EoR the fluctuations in 21-cm are mainly driven by the
fluctuations in $T_{\rm S}$, which is mildly correlated to the density
field, thus the redshift space distortions do not change the amplitude
and shape of the power spectrum drastically here{\footnote{However, one should
be cautious here while drawing any generic conclusions from these
results as they may be dependent on the assumptions that goes into the
models for the sources of heating.}}. They have also estimated the
$\mu$-decomposed power spectrum (right panel of Figure
\ref{fig:ps_cd}) following eq. \eqref{eq:rsd_mu} to further understand
the effect of redshift space distortions on the signal power
spectrum. They find that both the coefficients of $\mu^2$ and $\mu^4$
have non-zero values during this stage, however their amplitude are
smaller or comparable but not higher than the coefficient of $\mu^0$
term (which is equivalent to the real space power spectrum) in the
power spectrum. In a later work, \citet{ghara15} have computed the
anisotropy ratio (top panels of Figure \ref{fig:ani_r}) defined by
eq. \eqref{eq:r_mu} as a function of $z$ at large and intermediate
length scales ($k = 0.1$ and $0.5\,{\rm Mpc}^{-1}$) for the similar
reionization scenarios as in \citet{ghara14} but here they have also
included the sources of mass lower than $10^9\, {\rm M_{\odot}}$. It
is evident from these figures that the anisotropy due to the redshift
space distortions can be significant ($r_{\mu} \geq 1$) at large and
intermediate length scales even during the Cosmic Dawn and the early
stages of the EoR when one takes into account the inhomogeneities in
the X-ray heating and Lyman-$\alpha$ coupling.
\section{Quantifying the light cone effect in the 21-cm power
  spectrum}
\label{sec:quantifying_lc}
\subsection{During the EoR}
\begin{figure*}
\includegraphics[width=0.7\textwidth,angle=-90]{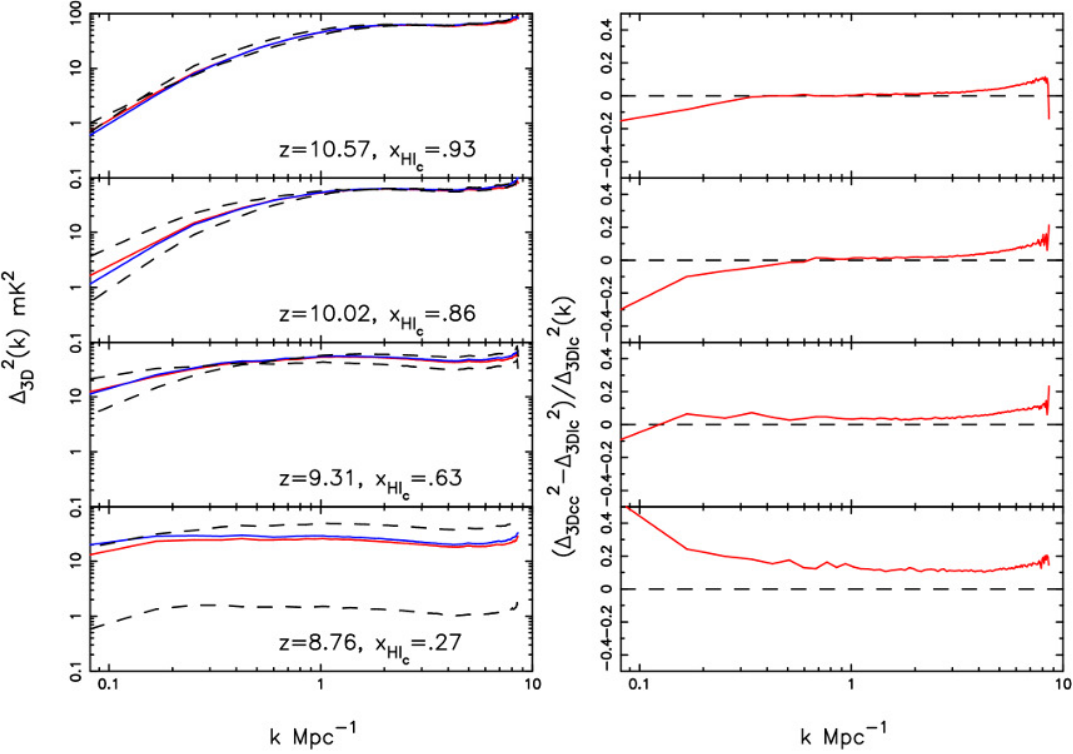}
\caption{The left panels show the spherically averaged 21-cm power
  spectrum estimated from the light cone cube (LC:red solid line) and
  coeval cube at the central redshift (CC:blue solid line) at
  different stages of the EoR. The dashed lines show the power
  spectrum from the coeval cubes at redshifts corresponding to the
  back and the front sides of the light cone box. The right panels
  show the relative difference in the power spectra between the LC
  and the CC boxes, estimated at the same stages of the EoR as shown on
  the left. Figure taken from \citet{datta12}.}
\label{fig:ps_lc_dif}
\end{figure*}
The first numerical investigation of the light cone effect on the
21-cm power spectrum from the EoR was done by \citet{datta12}. They
used a set of radiative transfer simulations of volume $(163\, {\rm
  Mpc})^3$ having three different reionization histories (from rapid
to slow) to study the light cone effect. They have quantified the
impact of the light cone effect by estimating the relative difference
between the spherically averaged power spectrum of the light cone box
[$\Delta^2_{{\rm LC}}(k)$] and the same estimated from a coeval cube
[$\Delta^2_{{\rm CC}}(k)$], corresponding to the central redshift of
the light cone box, {\it i.e.} through the quantity $[\Delta^2_{{\rm
    CC}}(k) - \Delta^2_{{\rm LC}}(k)]/\Delta^2_{{\rm LC}}(k)$ (Figure
\ref{fig:ps_lc_dif}). They observed that the relative change in the
power spectrum can be up to $\approx 50\%$ within the $k$ range $\sim
0.1-9.0\,{\rm Mpc}^{-1}$ and the large scales are affected more
compared to the small scales. They find that the light cone power
spectrum get enhanced at large scales and suppressed at small scale
compared to the coeval box power spectrum at the redshift of the
centre of the box. This enhancement and suppression of power happens
around a Fourier mode $k_{\rm cross-over}$, which gradually shifts
towards the large scales as reionization progresses. Using simple toy
models of power spectra, they argue that this behaviour is a signature
of the difference in the bubble size distribution in the coeval box
and the light cone box. They also observed that the difference in
between $\Delta^2_{{\rm LC}}$ and $\Delta^2_{{\rm CC}}$ is minimum
when reionization is half way through. The line of sight extent of the
light cone box influences the difference between $\Delta^2_{{\rm LC}}$
and $\Delta^2_{{\rm CC}}$. This difference goes down with the
reduction of the LoS extent of the light cone box. It was also noticed
that as any radio interferometric observation do not measure the
$k_{\perp} = 0$ mode, thus the DC value of the signal power spectrum
cannot be measured, which in turn removes the effect of the evolution
of mass averaged neutral fraction ($\xb$) (across the LoS) from the
observed light cone power spectrum. \citet{datta12} have further
observed that the change in reionization history (whether rapid or
slow), does not change the light cone affect on the power spectrum
significantly. The light cone effect is more dependent on the mass
averaged neutral fraction at the central redshift of the light cone
box rather than the total EoR history. These findings of
\citet{datta12} regarding the light cone effect were later confirmed
independently by \citet{laplante14} as well.

It is intriguing to ask the question, whether the light cone effect
introduces any significant LoS anisotropy to the observed EoR 21-cm
power spectrum{\footnote{As an alternative approach, one may first
    perform tomographic imaging on a smaller patch of the sky using
    the SKA deep survey and possibly find a functional form for the
    evolution of the signal which may enable us to correct for this
    effect in the power spectrum for shallow and medium
    surveys. However, one should also be cautious about the
    roboustness of such an approch. As the presence of an extremely
    bright source in the deep survey FoV will bias the fitting
    function for the light cone evolution of the signal which will
    lead to an erroneous correction of light cone effect in the medium
    and shallow survey. Also, as the signal is not detected yet,
    possibly one would first perform the shallow and medium surveys
    before going for deeper surveys.}}. If the anisotropy introduced
by the light cone effect is significant and also contributes in a
similar fashion in the power spectrum as the redshift space
anisotropy, then it would be really difficult to distinguish between
these two effects from a real observation. Also, several other
proposed outcomes using the SKA-LOW observations, which uses the
redshift space anisotropy as a tool, such as separating the cosmology
from astrophysics \citep{barkana05} or extracting the reionization
history \citep{majumdar16}, would be difficult to
achieve. \citet{datta14} have tried to address mainly this issue, {\it
  i.e.} whether the light cone effect introduces any significant
anisotropy to the signal power spectrum and what could be the best
observational strategy to minimize the impact of the light cone effect
on the EoR 21-cm power spectrum.

\citet{datta14} used a significantly large simulation volume ($[607
\,{\rm Mpc}]^3$), which is comparable to the field of view of LOFAR,
to study the impact of the light cone effect on the signal. The large
simulation volume allowed them to study this effect in case of both
very rapid as well rather slow reionization histories. They find the
impact of the light cone effect on the spherically averaged power
spectrum is maximum when the reionization is $\sim 20\%$ and $\sim
80\%$ finished and it is rather small at $\sim 50\%$
reionization. Which, reconfirms the findings of
\citet{datta12}. However, they do not observe any significant LoS
anisotropy introduced to the power spectrum due to the light cone
effect. They used a toy model to explain this rather surprising
finding. They find that, even though the light cone effect makes the
\HII bubbles larger in an observational data cube towards the side
closer to the observer compared to the opposite side of the
observational volume along the LoS, it does not necessarily makes the
\HII bubbles elongated or compressed along the LoS. This systematic
change in \HII bubble size along the LoS does not make the power
spectrum anisotropic as long as the bubbles remain approximately
spherical in shape. The power spectrum can only become anisotropic
when the shape of the individual \HII bubbles distort along the LoS,
which can only happen if the major sources of reionization are
extremely strong in terms of their photon emission rate ({\it e.g.}
quasars), which will lead to a relativistic growth of the \HII
regions. They also studied the power spectrum by including both
redshift space distortions and light cone effect to the signal and
found that the light cone effect does not change the signatures of the
redshift space distortions at any stage of the EoR.

\citet{datta14} also identified that there exists an optimal frequency
bandwidth along the LoS for the power spectrum estimation, within
which the light cone effect can be ignored without loosing the signal
to the uncertainties due to the sample variance. They found that for
large scales such as $k = 0.16\,{\rm Mpc}^{-1}$ this is $\sim 11$ MHz
and for intermediate scales $k = 0.41\,{\rm Mpc}^{-1}$ this is $\sim
16$ MHz, if one allows a change of $10\%$ in the power spectra. These
optimal bandwidths may change depending on the reionization history.
\subsection{During the Cosmic Dawn}
\begin{figure*}
\includegraphics[width=1.0\textwidth,angle=0]{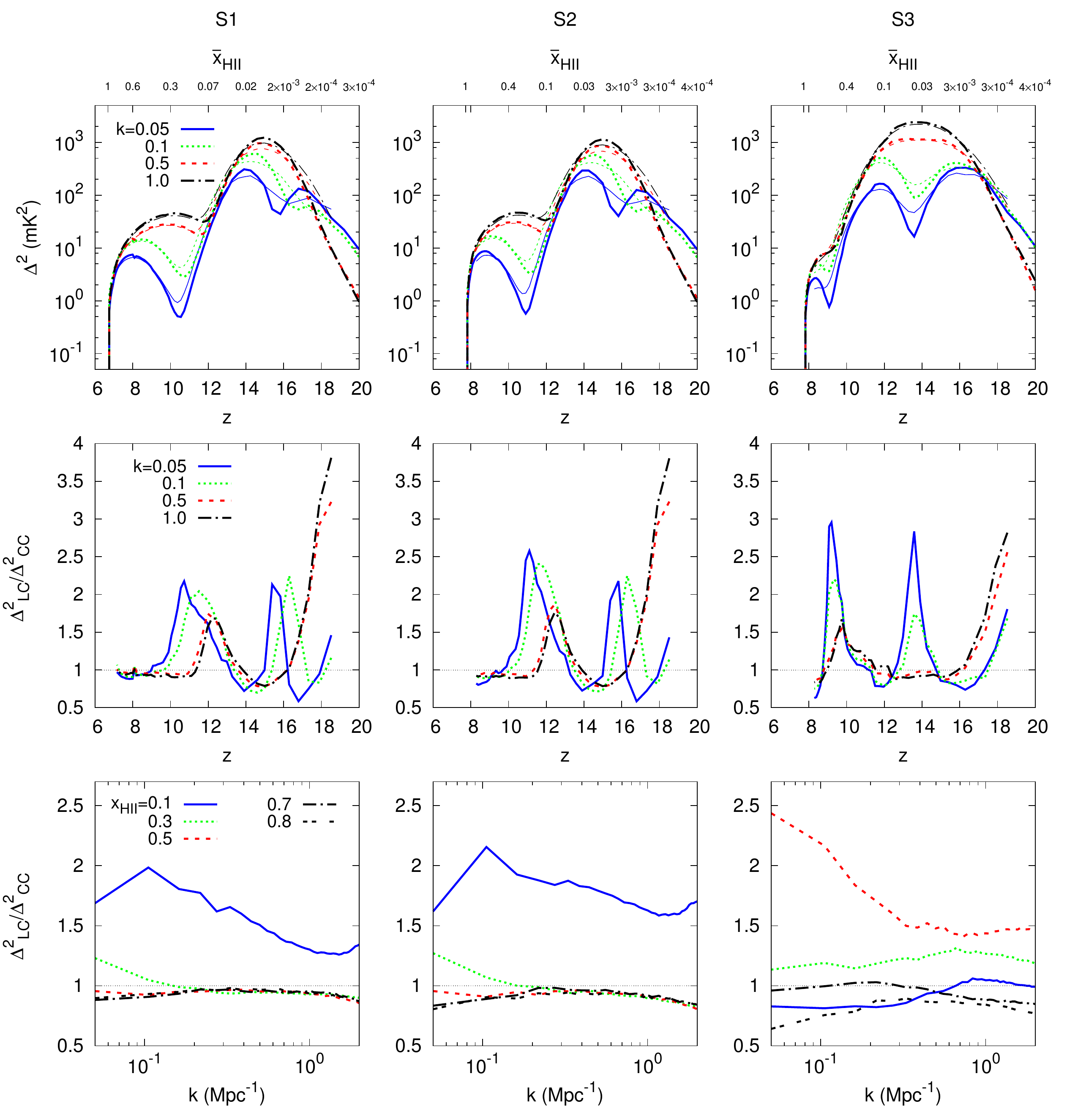}
\caption{The top panels show evolution of the spherically averaged
  21-cm power spectrum with redshift from simulation boxes with (thin
  line) and without (thick line) light cone effect in them. The middle
  and the bottom panels show relative change in the power spectrum due
  to the light cone effect when compared with the coeval power
  spectra. Three different columns correspond to three different
  models for reionization sources. Figure taken from \citet{ghara15}.}
\label{fig:ps_lc_cd}
\end{figure*}
\citet{ghara15} performed the first numerical study on the impact of
the light cone effect on the redshifted 21-cm power spectrum from the
Cosmic Dawn. They used a set of one dimensional radiative transfer
simulations (similar to \citealt{ghara14}) for the signal, while
considering spin temperature fluctuations in the signal due to the
inhomogeneous X-ray heating by the first sources and non-uniform
Lyman-$\alpha$ coupling. They found that the impact of the light cone
effect is more dramatic when one considers both the inhomogeneous
X-ray heating and the Lyman-$\alpha$ coupling induced spin temperature
fluctuations in the signal, compared to the case when these effects
are ignored. It was observed that at large length scales $k \sim
0.05\,{\rm Mpc}^{-1}$, light cone effect is more prominent around the
peaks and dips of the power spectrum when plotted as a function of
redshift (Figure \ref{fig:ps_lc_cd}). They observed that the large
scale signal power spectrum is suppressed around the three distinct
peaks (by factors of $\approx 0.7$) and enhanced (by factors of
$\approx 2$) around the dips due to this effect. This
enhancement/suppression was found to be higher in a case where one
includes sources of mass lower than $10^9\,{\rm M_{\odot}}$ in their
reionization prescription. A significant light cone effect was also
observed at small length scales ($k \sim 1\,{\rm Mpc}^{-1}$), during
the Cosmic Dawn.

The reason behind this behaviour is that where ever the power spectrum
experiences any non-linear evolution, the light cone effect becomes
substantial at those points, as any linear evolution will be mostly
cancelled out \citep{datta12}. This is possibly why this effect appears
to be more prominent around the heating peak and the peak and dip
caused due to the inhomogeneous Lyman-$\alpha$ coupling. It was also
observed that inclusion or exclusion of light cone effect changes the
power spectrum at large scales ($k \sim 0.05\,{\rm Mpc}^{-1}$) by
$-100$ to $100\,{\rm mK}^2$ and at small scales ($k \geq 0.5\,{\rm
  Mpc}^{-1}$) by $-250$ to $100\,{\rm mK}^2$. These large differences
should be in principle detectable by SKA-LOW, due to its higher
sensitivity. It was proposed that the power spectrum peaks at large
scales during the CD can be used to extract the source properties,
X-ray and Lyman-$\alpha$ backgrounds etc \citep{mesinger14}. However,
while performing such an exercise on an actual observational data set,
one needs to take into account the light cone effect as it suppresses
those peaks.

Similar to \citet{datta14} in the EoR, \citet{ghara15} also did not
find any significant LoS anisotropy introduced by the light cone
effect in the signal from the CD. They had estimated the anisotropy
ratio $r_{\mu}$ for large and intermediate scales (Figure
\ref{fig:ani_r}) to quantify the LoS anisotropy due to the light cone
effect. They found that the LoS anisotropy quantified by this ratio is
less than $5\%$ during most of the duration of the CD and the
EoR. They had also observed that, the light cone anisotropy do not
destroy the anisotropy due to the redshift space distortions present
in the signal.
%----------------------------------------------------------------------
\section{Detectability of the line of sight anisotropy signatures}
\label{sec:detectability}
\subsection{Foreground avoidance versus foreground removal}
One of the major obstacle for the detection of the CD-EoR 21-cm signal
is the foreground emission from galactic and extra-galactic point
sources, which is expected to be few orders of magnitude larger than
the signal itself. These foregrounds are assumed to be spectrally
smooth, which implies that they will only affect the lowest $k$ modes
parallel to the LoS ($k_{\parallel}$). However, due to the frequency
dependence of an interferometers's response, the foregrounds will
propagate into higher $k$ modes. This causes the foregrounds to become
confined to a wedge-shaped region in $k_{\parallel}-k_{\perp}$ plane,
which was first identified by \citet{datta10}. One of the ways to deal
with the foregrounds is thus to avoid this region where foregrounds
will be dominant and restrict the signal power spectrum estimation
within the region of the $k_{\parallel}-k_{\perp}$ space which is
expected to be clean from foregrounds (see {\it e.g.}
\citealt{trott12,dillon14,pober14} etc){\footnote{In the discussion of
    the wedge, one should keep in mind that the `chromatic sidelobes'
    are confined to the wedge, but the foregrounds themselves are not
    necessarily. When one Fourier transforms a declining function
    (e.g. spectrum of a source) there might be power also above the
    wedge. So a proper foreground subtraction is probably a
    pre-requisite for a believable result.}}. One drawback of this
method is that one have to live with the low SNR of the signal power
spectrum, after throwing away a large fraction of the data in this
way. The other drawback will be that the directional dependence of the
power spectrum will be hard to quantify in this method. Detectability
of any LoS anisotropy (more precisely the anisotropy due to the
redshift space distortions in this context) present in the 21-cm
signal power spectrum will depend on the fact that the power spectrum
is sampled uniformly over all angles (or $\mu$ values) with respect to
the LoS. Any sort of biased sampling of the power spectrum in this
regard may lead to misinterpretation of signal characteristics and the
related characteristics of the CD-EoR parameters. This would be
inevitable in this foreground avoidence technique
\citep{pober15}. Even when one is interested only in the spherically
averaged power spectrum, it has been observed that the biased sampling
of the $k_{\parallel}-k_{\perp}$ plane will introduce an atrificial
but significantly large bias to the estimated power spectrum
\citep{jensen16}. Thus to quantify the LoS anisotropies in the CD-EoR
21-cm signal using the future SKA-LOW shallow or medium survey
\citep{koopmans15}, it is very important that we employ a proper
foreground subtraction technique rather than using foreground
avoidance.
\subsection{Uncertainties due to the cosmic variance}
\label{subsec:cosmic_variance}
Any statistical estimation of the CD-EoR 21-cm power spectrum comes
with an intrinsic uncertainty of its own, which arises due to the
uncertainties in the signal across its different statistically
independent realizations, {\it i.e.} due to the cosmic variance of the
signal. In most of the analysis present in the current literature
related to the characterization of the CD-EoR 21-cm signal power
spectrum, it has been assumed that the signal has properties similar
to a Gaussian random field, which makes its cosmic variance to scale
as the square root of the number of independent measurements. This
could be a reasonably good assumption during the early phases of
reionization, but during the later stages of the EoR, the signal
becomes highly non-Gaussian as it gets characterized by the \HII
regions around the reionization sources
\citep{bharadwaj05,bharadwaj05a}. The size and population of these
\HII regions gradually grow as reionization progresses, making the
signal more and more non-Gaussian. A similar picture can be drawn
during the early stages of the cosmic dawn as well, when the signal is
characterized by the heated regions around the first sources of
light. Using a large ensemble of simulated EoR 21-cm signal,
\citet{mondal15} was the first to study the impact of the
non-Gaussianity of this signal on the cosmic variance of its power
spectrum estimation. They had shown that for a fixed observational
volume it is not possible to obtain an SNR above a certain limit
[SNR]$_l$, even when one increases the number of Fourier modes for the
estimation of the power spectrum. This limiting SNR stays
approximately in the range [SNR]$_l \sim 500$ to $10$, if we only
consider the effect of cosmic variance. The non-Gaussianity in the
signal increases as reionization progresses, and [SNR]$_l$ falls from
$\sim 500$ at $\xb = 0.9$ to $\sim 10$ at $\xb = 0.15$ for the [150
  Mpc]$^3$ simulation volume. It is possible to increase the SNR by
increasing the observational (or simulation) volume. In a more
detailed follow up work, \citet{mondal16} had provided an theoretical
framework to interpret the entire error covariance matrix of the
signal power spectrum. They identify two sources of contribution in
the error covariance. One is the usual variance of Gaussian random
field and other is the trispectrum of the signal, which comes due to
the fact that the different Fourier modes in the signal are correlated
due their inherent non-Gaissianity. They establish the fact that
errors in differnt length scales of the EoR 21-cm power spectrum are
correlated.  In a further follow up on this, \citet{mondal16b} studied
the evolution of these errors with the evolving IGM neutral
fraction. Using the EoR simulations they had established that for any
mass averaged neutral fraction $\xb \leq 0.8$ the error variance will
have a significantly large contribution from the trispectrum of the
signal for any Fourier mode $k \geq 0.5\, {\rm Mpc^{-1}}$. This
dependence may change depending on the reionization source model and
the resulting 21-cm topology. It is important to properly quantify the
actual uncertainties present in the power spectrum of the EoR 21-cm
signal due to the cosmic variance as that will decide the statistical
significance with which the LoS anisotropies in the power spectrum can
be quantified.
%-----------------------------------------------------------------------
\section{Summary and Future Scopes with the SKA-LOW}
\label{sec:summary}

In most part of this review, we have been trying to stress on the fact
that the understanding of the LoS anisotropy in the redshifted 21-cm
is not only important for the current ongoing surveys of the CD-EoR
but it will be specifically very crucial for the future SKA-LOW
surveys of this era, as the SKA-LOW is expected to measure this signal
with an unprecedented sensitivity in both spatial as well as frequency
direction compared to any of its predecessors. In context of the
proposed survey strategies for the SKA-LOW \citep{koopmans15}, the
following issues related to LoS anisotropy in the signal would
be particularly important ---

\begin{itemize} 

\item It has been proposed that using the shallow survey (observing
  $10,000\, {\rm deg^2}$ in the sky for $10$ hr) with SKA-LOW one
  would be able constrain different parameters for the CD-EoR 21-cm
  signal by comparing it with a large ensemble of simulated 21-cm
  power spectra \citep{greig15}. However, as it has been shown in the
  previous few sections, that a proper accounting of the impact of the
  redshift space distortion, the light cone effect and the bias in the
  sampling of the $k_{\perp} - k_{\parallel}$ space would be necessary
  to get a unbiased estimation of these parameters. It would be
  important to take into account the AP effect as well to account for
  any anisotropies in the signal due to the non-Euclidean geometry of
  the space-time. Finally, the effect of the non-Gaussian nature of
  the signal on it's power spectrum error covariance needs to be taken
  into account. This will help to quantify the uncertainties in the
  estimated reionization parameters properly.

\item While estimating power spectrum from the observed three
  dimensional data one should estimate it within an optimal band width
  to avoid any significant impact due to the light cone effect. The
  width of this optimal bandwidth needs to be properly examined with a
  larger variety of reionization sources and reionization histories.

\item Accounting for the possible light cone effect would be also
  important, when one will try to characterize the heating sources
  during the CD from the peaks of 21-cm power spectrum, as this effect
  tends to reduce those peaks.

\item If the foreground removal works reasonably well, it would be
  possible to constrain the reionization history from the evolution of
  the quadrupole moment of the power spectrum estimated from the
  proposed medium deep survey by the SKA-LOW (observing $1000\, {\rm
    deg^2}$ of the sky for $100$ hr).

\item Presence of the noise bias and several telescope related
  anomalies in the observed data may make the quantification of the
  signal and the LoS anisotropy present in it very difficult. Thus it
  would be necessary to develop clever estimators of the signal power
  spectrum, which will inherently remove such bias and anomalies
  \citep{datta07a,choudhuri14,choudhuri16}.

\end{itemize}

%----------------------------------------------------------------------

\section*{Acknowledgments}
SM would like to acknowledge financial support from the European
Research Council under the ERC grant number 638743-FIRSTDAWN and from
the European Unions Seventh Framework Programme FP7-PEOPLE-2012-CIG
grant number 321933-21ALPHA.  KKD would like to thank University Grant
Commission (UGC), India for support through UGC-faculty recharge
scheme (UGC-FRP) vide ref. no. F.4-5(137-FRP)/2014(BSR).
%--------------------------------------------------------------------

\bibliographystyle{mnras} 
\bibliography{refs}

%--------------------------------------------------------------------

%\bsp

%--------------------------------------------------------------------

\label{lastpage}

\end{document}